\documentclass[doublecol]{epl2}
\usepackage{amsmath}
\usepackage{amsfonts}
\usepackage{amssymb}

\newcommand{\abs}[1]{\left| #1 \right| }
\newcommand{\avg}[1]{\left\langle #1 \right\rangle}

\begin{document}
\title{Interactions destroy dynamical localization with strong and weak chaos} 
%
\author{G. Gligori\'c\inst{1,2} \and J. D. Bodyfelt\inst{1} \and S. Flach\inst{1}}
\institute{
\inst{1} Max-Planck-Institut f\"ur Physik komplexer Systeme, N\"othnitzer Stra\ss e 38, D-01187 Dresden, Germany \\
\inst{2} Vin\v ca Institute of Nuclear Sciences, P.O. Box 522, 11001 Belgrade,Serbia
}
\abstract{
Bose-Einstein condensates loaded into kicked optical lattices can be treated as quantum kicked rotor systems. Noninteracting rotors show dynamical localization in momentum space. The experimentally tunable condensate interaction is included
in a qualitative Gross-Pitaevskii type model based on two-body interactions. We observe strong and weak chaos regimes of wave packet spreading in momentum space. In the intermediate strong chaos regime the condensate energy grows as $t^{1/2}$. In the asymptotic weak chaos case the growth crosses over into a $t^{1/3}$ law. The results do not depend on the details of the kicking. 
}
%
%
\maketitle
\section{Introduction}
There is much interest in quantum behavior of systems that are chaotic in the classical limit. The hallmark model used in many such studies is the kicked rotor \cite{casati_stochastic_1979, chirikov_universal_1979}, which was observed
to exhibit a parametric transition from integrable dynamical behavior to a fully chaotic one. In physical terms, this corresponds to an unbounded diffusive increase of the rotor energy in the chaotic regime, as opposed to energies bounded by invariant tori in the integrable case.  Once quantized, the kicked rotor disrespects the classical chaotic energy increase - a highly interesting phenomenon known as \textit{dynamical localization}, similar to Anderson localization in disordered tight-binding models \cite{fishman_chaos_1982,grempel_quantum_1984}. Dynamical localization has been observed in laboratory experiments with atoms in microwave fields \cite{bayfield_localization_1989,bluemel_dynamical_1991}, and in ultracold dilute gases condensed within kicked optical lattices \cite{ryu_high-order_2006,behinaein_exploring_2006,kanem_observation_2007}. In particular, the latter realization with a Bose-Einstein condensate, where the magnitude and the sign of the interactions between atoms can be tuned via Feshbach resonance by an additional magnetic field, has greatly opened opportunities to study dynamical systems in the presence of many-body interactions. In the mean-field approximation, these interactions can be modeled by adding a nonlinear cubic term in the Schr\"odinger equation, giving the well-known Gross-Pitaevskii equation. This can also be extended to the quantum kicked rotor model, resulting in its nonlinear generalization. Considerations of nonlinearity's effects on dynamical localization have initially shown that with strong enough nonlinearity, quantum localization is destroyed and subdiffusive spreading occurs \cite{shepelyansky_delocalization_1993, garcia-mata_delocalization_2009}, similar to results from nonlinear disordered media. The possibility to view dynamical localization as a counterpart to Anderson localization encourages further detailed numerical studies of this problem - particular motivation is found in recent studies \cite{lemarie_observation_2009,lemarie_critical_2010} whereby a quantum kicked rotor multiply kicked with three incommensurate frequencies mimics the three-dimensional Anderson metal-insulator transition. In this paper, we shall focus specifically on the application of a theoretical framework developed for spreading of waves in nonlinear disordered media \cite{flach_spreading_2010, krimer_statistics_2010,laptyeva_crossover_2010,bodyfelt_nonlinear_2011}. Our main aim is to establish spreading laws for a nonlinear version of the quantum kicked rotor. In the first section, we give the model description. Next, theoretical predictions of spreading regimes are discussed. Numerical results are then presented, followed by our conclusions.
\section{Model}
Experiments of quantum kicked rotor systems within Bose-Einstein condensates, where many-body interactions play a significant role, focus theoretical attention on dynamical localization in the presence of nonlinear interactions. In the mean-field approximation, dynamics of the kicked rotor can be modeled by the following form of the Gross-Pitaevskii equation
\begin{equation}
i \hbar \frac{\partial\psi}{\partial t} = - \frac{\hbar^2}{2 M} \frac{\partial^2 \psi}{\partial \theta^2} + \tilde{\beta}\abs{\psi}^2 \psi + {\bar k} \cos(\theta) \cdot \psi \sum_m \delta(t-mT). \label{eq:NLS}
\end{equation}
Here $\tilde{\beta}$ is the nonlinear strength, which is proportional to the tunable two-body scattering length of atoms in a BEC. $M$ is the mass of the atoms, $\bar k$ is the perturbative kick strength, and $T$ is the period of applied kicks.  Note that the analogy between an abstract rotor and the atomic wavefunctions is obtained when the atoms are loaded into a momentum eigenstate of the lattice with a Bloch wavenumber of zero - spatially homogeneous kicks will keep the Bloch wavenumber invariant, yet allow changes in the momentum (see below).

When the nonlinear strength is zero ($\tilde{\beta}=0$), the problem is reduced to a linear map. Because of the instant action of the perturbation, the evolution operator $\hat{U}$ of the linear system can be written as product of two noncommuting unitary operators
\begin{equation*} 
\psi(\theta, t+1) = \hat{U} \cdot \psi(\theta, t) = \hat{B}(k,\theta) \cdot \hat{G}(\tau / 2,\theta) \cdot \psi(\theta, t) 
\end{equation*}
The first corresponds to free rotation between two successive kicks,
\begin{equation*} 
\hat{G}(\tau / 2,\theta) = \exp \left( - i \frac{\tau}{2} \frac{\partial^2}{\partial \theta^2} \right), \quad \tau \equiv \frac{\hbar T}{M},
\end{equation*}
and the second describes the evolution over the kick,
\begin{equation*} 
\hat{B}(k, \theta) = \exp\left( - ik \cos \theta \right), \quad k \equiv \bar{k} /\hbar.
\end{equation*}

The solution $\psi(\theta,t)$ can be expanded in an angular momentum basis
\begin{equation}
\psi(\theta,t) = \frac{1}{\sqrt{2\pi}} \sum^{\infty}_{n=-\infty} A_n(t) e^{i n \theta} \label{eq:AngMomMap}
\end{equation}
where the coefficients $A_n(t)$ are Fourier coefficients of the time-dependent wave function $\psi(\theta,t)$. As a result of the action of the evolution operator $\hat{U}$ on $\psi(\theta,t)$ over one period $T$, the following map for the Fourier coefficients is obtained
\begin{equation}
A_n(t+1) = \sum_ m (-i)^{n-m} J_{n-m}(k) A_m(t) e^{-i \frac{\tau}{2} m^2} \label{eq:FourierMap}
\end{equation}
where $J_{n-m}(k)$ is a Bessel function of the first kind. Unlike the classical rotor, the behavior of the quantum one essentially depends on two parameters. The perturbation strength $k$ sets the effective number of unperturbed states covered by one kick, while $\tau$ relates the period of applied kicks $T$ to the natural frequency of the rotor, which is defined as $\omega=\hbar/2M$. By $\omega=2\pi/\tilde{T}$, the parameter $\tau$ can be written as $\tau=4\pi T/\tilde{T}$. 

Since the perturbation is time-periodic, Floquet theory can be applied \cite{izrailev_simple_1990}
\begin{equation}
\psi(\theta,t) = e^{-i\chi t} \phi_{\chi}(\theta,t), \, \phi_{\chi}(\theta,t+1) = \phi_{\chi}(\theta,t)  \label{eq:QuasiE}
\end{equation}
where $\phi_{\chi}(\theta,t)$ are $t$-periodic functions with period $T$ (Floquet states), and $\chi$ are the quasienergies. Then Eq.(\ref{eq:FourierMap}) can be treated as an eigenvalue problem
\begin{equation}
\lambda_\nu A^{\nu}_n=\sum_m (-i)^{n-m} J_{n-m}(k) e^{-i \frac{\tau}{2}m^2} A^{\nu}_m \label{eq:floquet}
\end{equation}
with eigenvalues located on the unit circle in the complex plane: $\lambda_\nu=e^{i\chi_\nu}$. For a rational ratio of $T/\tilde{T}$, extended eigenvectors are obtained. This case corresponds to the regime of quantum resonances, where the rotor energy $E(t)= \sum_n \abs{A_n}^2 n^2 / 2$ grows as $t^2$ (ballistically) in time - the cause being quantum interference effects \cite{casati_stochastic_1979}. For an irrational ratio of $T/\tilde{T}$ (the case treated here), the sequence $\lbrace e^{-i \frac{\tau}{2} m^2} \vert \, m\in \mathbb{Z} \rbrace$ is quasiperiodic and the corresponding eigenvectors are exponentially localized - this effect is \textit{dynamical localization}. If the argument $\tau m^2 / 2$ in the exponential is replaced by a truly uniform random sequence in the interval $\left[0,2\pi\right]$, calculations have shown the same effect of exponential localization in corresponding eigenvectors, supporting the idea that quasiperiodic sequences mimic disorder \cite{brenner_pseudo-randomness_1992}.     

For a given set of parameters $\tau$ and $k$, the localization extent of an eigenvector is estimated by its participation number: $P = 1/\sum_ n \abs{A^{\nu}_n}^4$. We define the localization volume as the average of this over the mode index, $V=\avg{P}_\nu$. This localization volume depends on the strength of kick $k$, and is increasing with $k$. It measures the typical size of a localized eigenvector. The parameter $\tau$ has no significant influence on the localization volume, provided that the ratio $\tau/4\pi$ is irrational. The localization volume also defines the number of other modes a given eigenvector can interact with, once the nonlinearity is added. Since all these $V$ modes have in general different quasienergies $\chi_{\nu}$, we can introduce a characteristic difference scale - the average quasienergy spacing $d$. Since all quasienergies are modulo $2\pi$, it is simply given by 
\begin{equation}
d = 2\pi / V\;.
\label{spacing}
\end{equation}      

In the nonlinear case ($\tilde{\beta}\neq 0$), expanding $\psi(\theta,t)$ in the angular momentum basis, Eq.(\ref{eq:AngMomMap}), the dynamics between two successive kicks is described by
\begin{equation}
i\frac{\partial A_n}{\partial t} = - \frac{1}{2}\tau n^{2} A_n + \beta \sum_{n_{1}}\sum_{n_{2}} A^{*}_{n_{1}} A_{n_{2}} A_{n-(n_{2}-n_{1})}, \label{eq:momNLS} 
\end{equation}
where $\beta = \tilde{\beta}T/2\pi\hbar$. Since we are ultimately interested in the limit of small amplitudes, we treat the nonlinear term perturbatively and keep only the diagonal terms in Eq.(\ref{eq:momNLS}). Neglecting all off-diagonal terms in Eq.(\ref{eq:momNLS}) and integrating over the free motion between two delta kicks, $A_{n}(t)$ evolves according to
\begin{equation}
A_{n}(t+1) = A_{n}(t) e^{-i\frac{\tau}{2}n^2 + i\beta\abs{A_n}^2}, \label{eq:momBK} 
\end{equation}
After additional integration over the infinitesimal interval of one kick, the map - which now describes the evolution over one whole period - becomes
\begin{equation}   
A_n(t+1)=\sum_m (-i)^{n-m} J_{n-m}(k) A_m(t) e^{-i\frac{\tau}{2}m^2 + i\beta\abs{A_m}^2}. \label{eq:NLS_Shep}
\end{equation}
This map was first introduced by Shepelyansky in \cite{shepelyansky_delocalization_1993}. Comparison of the results of this map with direct numerical simulation of the corresponding model, Eq.(\ref{eq:NLS}), has shown differences on a short time scale, but the same asymptotic behavior in the rotor energy \cite{rebuzzini_delocalized_2005}. At the same time, this model allows for more efficient and faster numerical computation. 

Numerical simulations of the map, Eq.(\ref{eq:NLS_Shep}), are performed for different parameter sets $\lbrace k, \tau, \beta\rbrace$, starting with a single-site excitation at site $n_0$: $\abs{A_n(t)}^2 = \delta_{n,n_0}$. The lattice sizes were between $N=2^{10}$ for smaller $k$ values and $N=2^{11}$ for larger values. To characterize the time evolution properties, we compute the on-site probability $w_n=\abs{A_n}^2$, the time-dependent participation numbers $P=1/\sum_n w_n^2$, and the average energy. Since localization is in momentum space, the average energy is defined by the second moment of the on-site probability, multiplied with one half 
\begin{equation}
E = \sum_n \frac{1}{2} \left( n-\bar{n} \right)^2 w_n, \label{eq:avgerg}
\end{equation}
where $\bar{n}=\sum_n n \, w_n$ is the first moment. To quantify the sparsity of a packet we calculate the compactness index as a ratio of the square of the participation number and the second moment, $\zeta=P^{2}/2E$ \cite{skokos_delocalization_2009}. To validate our numerics, we ensure the total norm, $S=\sum_n w_n$, is conserved (at absolute relative errors of $\le 10^{-8}$) during all simulations.
\section{Regimes of Spreading}
In Eq.(\ref{eq:NLS_Shep}), the nonlinearity induces a quasienergy shift of $\Delta\phi_m=\beta\abs{A_m}^2$. It also induces an interaction between localized eigenvectors. Namely, the nonlinear term of the map in Eq.(\ref{eq:NLS_Shep}) can be rewritten as $\exp(i\beta\abs{A_m}^2)=1+f(i\beta\abs{A_m(t)}^2)$. Expanding $A_n$ in the linear eigenbasis, $A_n(t)=\sum_\nu \phi_\nu(t) A^{\nu}_n$, Eq.(\ref{eq:NLS_Shep}) is transformed to
\begin{eqnarray*}
\phi_\nu(t+1) &=& \lambda_\nu \phi_\nu(t) + \sum_{n, \mu, m} \phi_\mu(t) \cdot (-i)^{n-m} J_{n-m}(k)  \cdot \\ 
& & e^{-i\frac{\tau}{2} m^2} A^{\nu*}_n A^{\mu}_m \cdot f\left(i \beta \abs{\sum_{\mu '} \phi_{\mu '}(t) A^{\mu '}_m}^2\right) 
\end{eqnarray*}
This is the corresponding map in the basis of localized complex eigenmodes, $\phi_\nu(t)$. Under small quasienergy shifts, the nonlinear term is Taylor expanded ($f \approx i \beta \abs{A_m(t)}^2$) to yield
\begin{equation}
\phi_\nu (t+1) = \lambda_\nu \phi_\nu(t) + \beta \sum_{\mu_{1}, \mu_{2}, \mu_{3}} I_{\nu, \mu_{1}, \mu_{2}, \mu_{3}} \phi_{\mu_{1}} \phi^{*}_{\mu_{2}} \phi_{\mu_{3}} \label{eq:NMS}
\end{equation}
with the overlap integral
\begin{eqnarray}
I_{\nu,\mu_{1},\mu_{2},\mu_{3}} &=& \sum_{n,m}(-i)^{n-m+1}J_{n-m}(k) \cdot \nonumber \\
& & e^{-i \frac{\tau}{2}m^{2}} A^{\nu *}_{n} A^{\mu_{1}}_{m} A^{\mu_{2}*}_{m} A^{\mu_{3}}_{m}  \label{eq:overlapI}
\end{eqnarray}
Since all eigenvectors are exponentially localized, each is effectively coupled with a finite number of neighbors - the interaction has finite range. However, according to norm conservation, as the wavepacket spreads, individual on-site probabilities will start to reduce - so will the average norm density, $n$. Since the coupling strength between eigenmodes is proportional to $n$, it also decreases with packet spreading, while the number of excited eigenmodes grows.

In disordered nonlinear wave equations, spreading regimes were observed \cite{laptyeva_crossover_2010,bodyfelt_nonlinear_2011} to depend on relations between a nonlinear-induced frequency shift, the linear spectral width, and the average frequency spacing between interacting eigenmodes. In the case of frequency shift larger than the linear spectral width, at least a part of the wavepacket is self-trapped \cite{kopidakis_absence_2008,skokos_delocalization_2009}. In weaker cases, when self-trapping is avoided, two outcomes are possible. If the frequency shift is larger than the average spacing, almost all eigenmodes resonantly interact with each other - the \textit{strong chaos} regime. Otherwise if the frequency shift is less than the average spacing, there is very minimal resonant interaction amongst eigenmodes - the \textit{weak chaos} regime. 

The similarity of the models hints that the spreading conjecture of \cite{flach_spreading_2010} may also be valid for the nonlinear quantum kicked rotor. In this conjecture, an exterior eigenmode within a localization volume of the wavepacket's boundary is incoherently heated by the packet. The probability of a resonance inside the packet depends on the ratio between the nonlinear-induced quasienergy shift, $\delta \approx \beta n$, and the average quasienergy spacing $d$: $\mathcal{P}(\beta n) \approx 1 - e^{-\beta n/d}$ \cite{krimer_statistics_2010, flach_spreading_2010}. From Eq.(\ref{eq:NMS}), it follows that the exterior eigenmode is excited according to
\begin{equation}
\phi_\nu (t+1) \approx \lambda_\nu \phi_\nu(t) + \beta n^{3/2} \mathcal{P}(\beta n) f(t), \label{eq:NMS1}
\end{equation}
where $f(t)$ is a stochastic force. The time-dependent exterior eigenmode norm is $\abs{\phi(t)}^{2} \sim \beta^{2}n^{3}(\mathcal{P}(\beta n))^{2}t$. It excites to the level of the packet norm density $n$ in a time $T \sim \beta^{-2}n^{-2}(\mathcal{P}(\beta n))^{-2}$. At this point, the once exterior eigenmode can be considered to be engulfed into the packet interior, so the rate of norm diffusion can be defined as $D = T^{-1} \sim \beta^{2}n^{2}(\mathcal{P}(\beta n))^{2}$. The second moment is inversely proportional to the square norm, $m_{2} \sim 1/n^{2}$ and the diffusion equation is $m_{2} \sim Dt$. From these two relations, one obtains an equation $1/n^{2} \sim \beta(1-e^{-\beta n/d})t^{1/2}$. Given the limits of the resonance probability, this yields two regimes of subdiffusive spreading
\begin{eqnarray*}
m_{2} &\sim& \beta t^{1/2},\ \ \ \ \beta n/d > 1,\ \ \text{strong chaos} \\
m_{2} &\sim& d^{-2/3}\beta^{4/3}t^{1/3},\ \ \beta n/d < 1,\ \ \text{weak chaos}
\end{eqnarray*}

The average energy of the quantum kicked rotor $E$ is defined by the second moment of the on-site probabilities, Eq.(\ref{eq:avgerg}). Hence, from the two above regimes, theoretical expectations for the subdiffusive growth of the average energy, $E \sim t^\alpha$, are $\alpha \approx 1/2$ in the strong chaos regime, and $\alpha \approx 1/3$ in the weak chaos regime. Spreading increases the wavepacket width, but under a norm conservation, individual on-site probabilities are reduced and the nonlinear quasienergy shift likewise decreases. A wavepacket launched in the strong chaos regime consequently spreads to a point whereby quasienergy shifts are reduced to only minimal resonant interactions. The strong chaos regime is therefore a transient one; eventually the dynamics will cross over to approach the weak chaos regime in an asymptotic limit. Nonetheless, the previous studies in disordered nonlinear Anderson or Klein-Gordon chain have shown that this transient regime can exist for many orders of magnitude before the crossover is observed \cite{laptyeva_crossover_2010,bodyfelt_nonlinear_2011}. 

\section{Numerical Results}                       
\begin{figure}[hbt]
\begin{center}
\includegraphics [width=\columnwidth,keepaspectratio,clip]{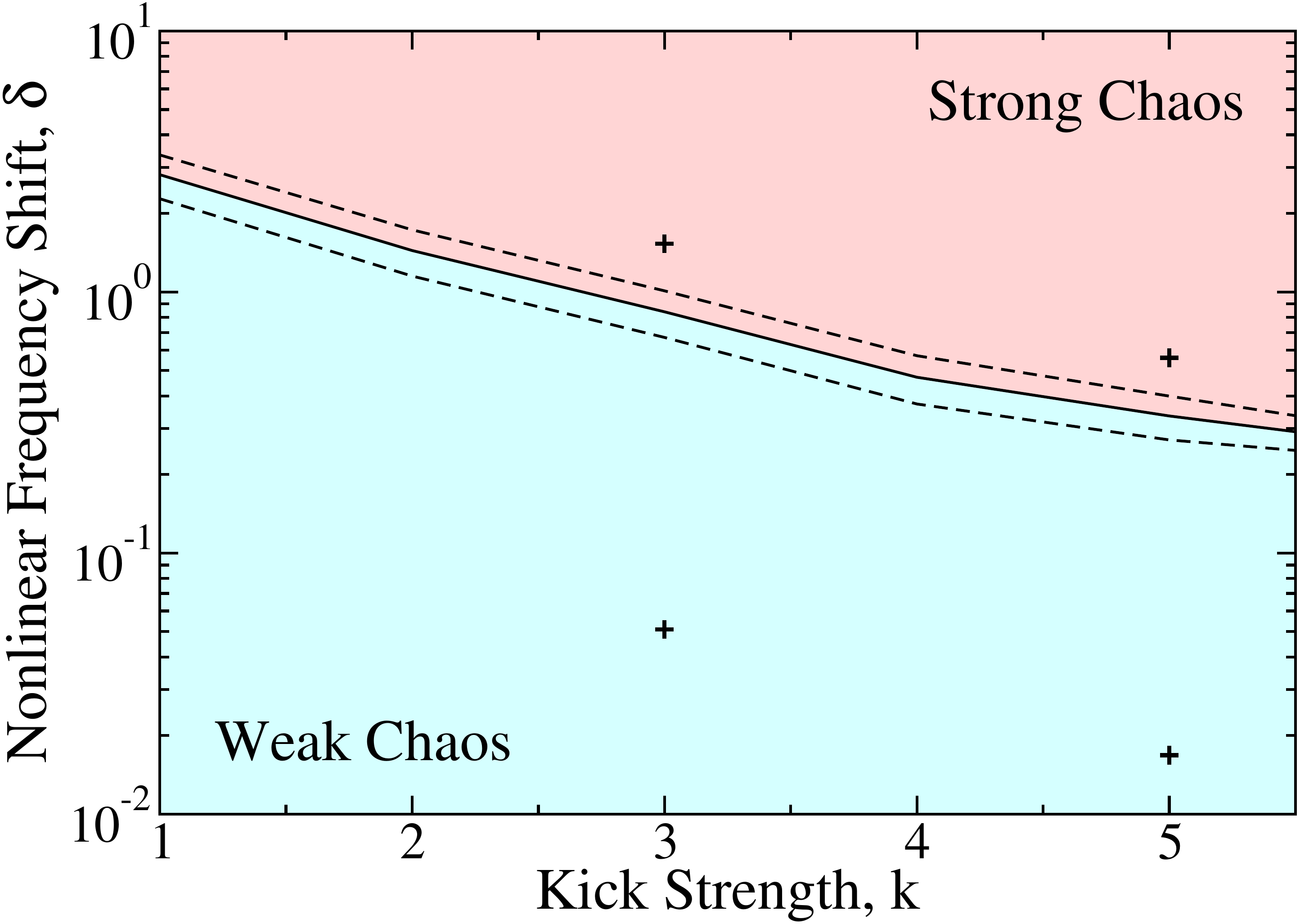} 
\caption{The parameter space as given by the nonlinear quasienergy shift $\delta$ versus the kick strength $k$. The seperation between the two expected regimes is estimated by $d$, the average spacing between quasienergies. The full line is $d$ for the quasiperiodic sequencing ($\tau=1$), while dashed lines correspond to its error within one standard deviation. Crosshairs correspond to parameters used for numerical simulations.} 
\label{fig:parameter_space}
\end{center}
\end{figure}

Since all eigenvalues are located on the unit circle for the quantum kicked rotor, there is essentially no limit to the linear spectral width. Therefore, unlike in disordered chains, the self-trapping regime is not expected to be observed, regardless of the nonlinearity strength. For irrational $\tau$, all quasienergies are basically distributed homogeneously over the interval $(0,2\pi)$. The localization volume, $V$, depends on the kick strength $k$. The average spacing between quasienergies can then be estimated using Eq.(\ref{spacing}). In Fig.\ref{fig:parameter_space}, the dependence of the average spacing $d$ on the kick strength $k$ is shown, for a quasiperiodic sequence ($\tau=1$). The dependence 
$d\left(k\right)$ for a random sequence is quite close (well inside the shown standard deviation) to that obtained for the quasiperiodic case, therefore it is not shown.

Starting from a single-site excitation, before quantum supression effects become significant, the packet will extend over the localization volume. Consequently, its norm density will drop from the initial value $n_{in}=1$ to $n\approx 1/V$. Choosing values of the nonlinear coefficient $\beta$ gives quasienergy shifts of $\delta \approx \beta n$ either above or below the line $d\left(k\right)$. Considering that the shift $\delta$ and the average quasienergy spacing $d$ are both inversely proportional to the localization volume, the condition $\delta>d\left(k\right)$ can be reduced to $\beta>2\pi$. This effectively gives us a parameter space in which to choose the nonlinearity parameter $\beta$. The weak chaos regime is expected for the values $\beta<2\pi$, while the strong chaos regime is expected for values $\beta>2\pi$. Simulations were performed for the values $k=3,5$ for both a quasiperiodic sequence ($\tau=1$) and for a random sequence. For each set of parameters, the simulations were over more than $500$ different realizations. For random sequences, these realizations were simply unique uncorrelated sequences, while for the quasiperiodic case the unique realizations were different initially excited states of the quantum kicked rotor. To suppress fluctuations, the logarithms (base 10) of time dependent measures are averaged over these realizations - we do this for the average energy in Eq.(\ref{eq:avgerg}) to give $\avg{\log_{10} E}$. In order to quantify our findings, we utilize locally weighted regression smoothing \cite{cleveland_locally_1988}, and then calculate the corresponding local derivative
\begin{equation}
\alpha=\frac{d \avg{\log_{10} E} }{d \left( \log_{10} t \right) } \label{eq:deriv}
\end{equation} 
via a central finite difference \cite{hoffman_numerical_1992}. 

\begin{figure}[hbt]
\begin{center}
\includegraphics[width=\columnwidth,keepaspectratio,clip]{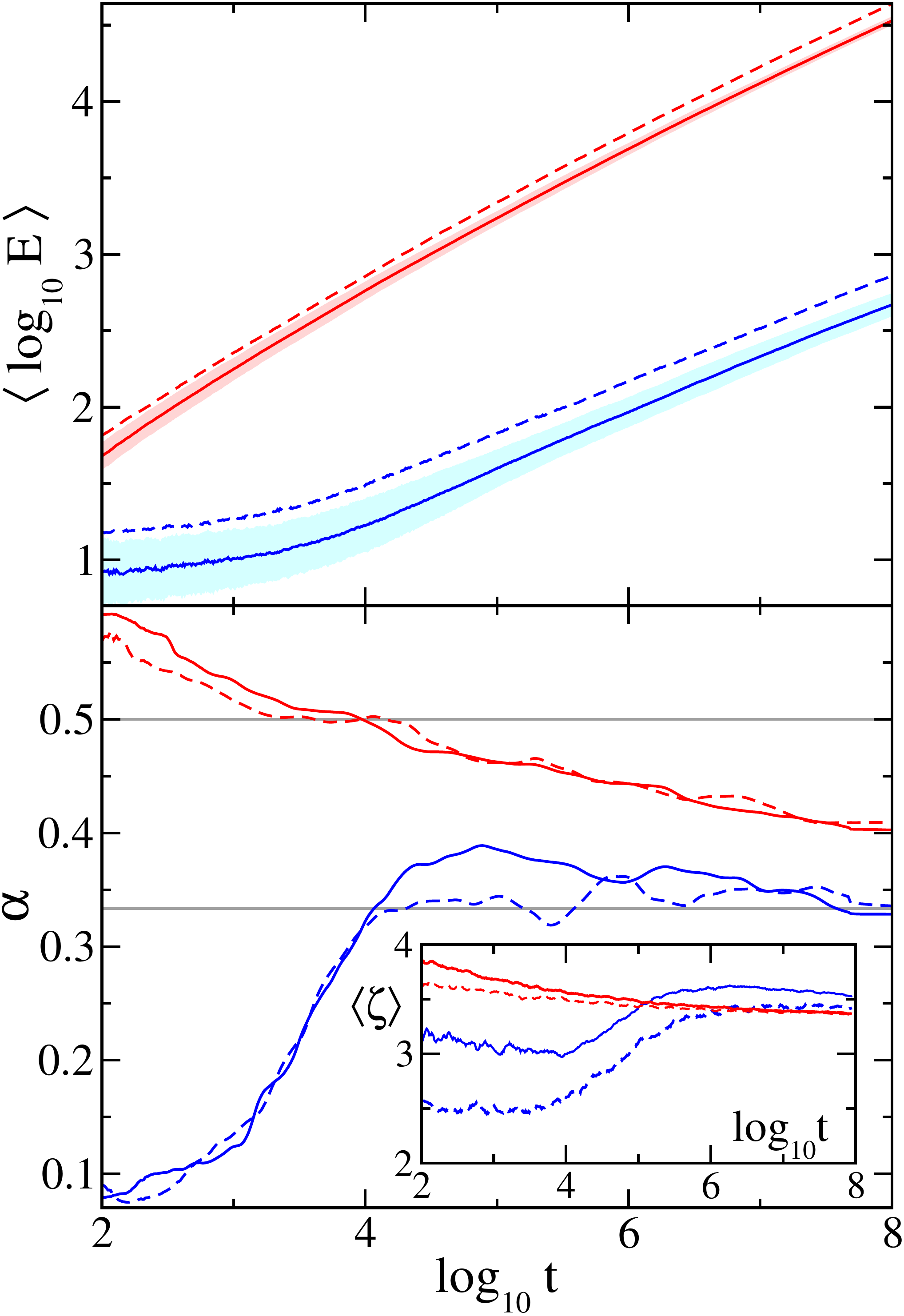}
\caption{Under a kick strength of $k=3$, measures for $\beta=0.4$ (blue) and $\beta=12$ (red), for both quasiperiodic sequences set by $\tau=1$ (solid line), and for random sequences (dashed line). Upper row: Mean logarithms for energy $\avg{\log_{10} E}$. The clouds around the quasiperiodic sequences correspond to one standard deviation error. Lower row: finite-difference derivative of the above. Grey horizontal lines correspond to exponents for weak and strong chaos regimes. Inset: average compactness index $\avg{\zeta}$ as a function of time.} 
\label{fig:k3} 
\end{center}
\end{figure}

For $k=3$, the results are seen in Fig.\ref{fig:k3}. In the weak chaos regime, $\beta=0.4$ (blue curves), the exponent $\alpha$ increases slightly above the expected $1/3$, but reaches it at later times. In the strong chaos regime, $\beta=12$ (red curves), the exponent $\alpha$ decreases from the predicted value $1/2$, without spending much time in the strong chaos regime. These findings are then compared with the results for random sequences, shown as dashed lines. As can be seen, spreading behaviors are quite similar to the quasiperiodic sequencing. Additionally in the inset of Fig.\ref{fig:k3}, the average compactness index $\avg{\zeta}$ asymptotically attains values in the range $\sim 3.4-3.6$ for both the quasiperiodic and random cases. Note that the compactness index does not significantly decay, meaning that subdiffusive growth of the second moment is followed by corresponding subdiffusive growth of the participation number. This indicates almost all states inside the packet are homogeneously populated during the spreading \cite{garcia-mata_delocalization_2009}. 

Lower values of the kick strength $k$ correspond to smaller localization volumes. These lower values may be associated with no clear identification of the strong chaos regime in the kicked rotor; since small localization volumes are related to a small interaction range between eigenmodes, it ought be that the strong chaos regime is quite short in time for lower values of the kick strength $k$. To test this argument, we performed simulations for $k=5$, but with the total number of kicks one order smaller than for $k=3$. The reason for this reduction is simply that bigger $k$ requires larger system sizes, increasing computational time.  

\begin{figure}[htb]
\begin{center}
\includegraphics[width=\columnwidth,keepaspectratio,clip]{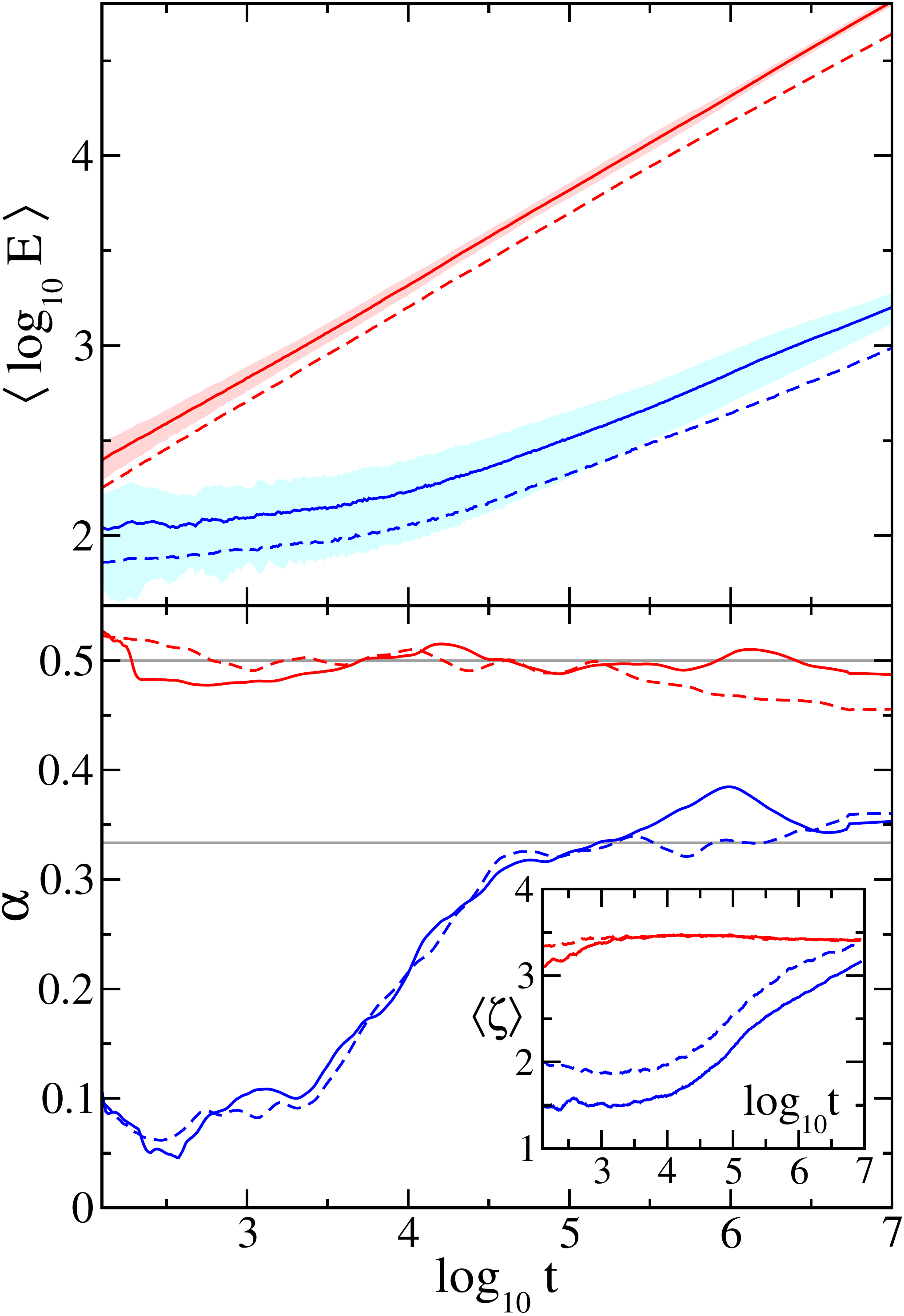}
\caption{Under a kick strength of $k=5$, measures for $\beta=0.3$ (blue) and $\beta=10$ (red), for both quasiperiodic sequences set by $\tau=1$ (solid line), and for random sequences (dashed line). Upper row: Mean logarithms for energy $\avg{\log_{10} E}$. The clouds around the quasiperiodic sequences correspond to one standard deviation error. Lower row: finite-difference derivative of the above. Grey horizontal lines correspond to exponents for weak and strong chaos regimes. Inset: average compactness index $\avg{\zeta}$ as a function of time.}  
\label{fig:k5}
\end{center}
\end{figure}

For $k=5$, the results are shown in Fig.\ref{fig:k5}. The derivative $\alpha$, for $\beta=0.3$ (blue curves), shows similar behavior to the case for $k=3$ in the weak chaos regime - namely, it grows until it slightly exceeds and then drops back to $1/3$. Even though for $k=5$ we are unable to look at extremely long times, we can still conclude that observed behavior is not much qualitatively different in the weak chaos regime than for the case $k=3$. On the other hand, for $\beta=10$ (red curves), which should correspond to the transient strong chaos regime, it is clear that the exponent $\alpha$ stays for a significantly long time near its theoretically predicted value of $1/2$. The compactness index shows the same behavior as in $k=3$, indicating homogeneity of the populated states within spreading  packets. The results for the random sequences (dashed lines) also confirm similar behaviors for $k=5$. In particular, at the end of the observed time, it can be seen that the exponent $\alpha$ decreases below $1/2$, signaling an entry of the system into a regime crossover from strong to weak chaos.
\section{Conclusion}\label{sec:conclude}
The results of extensive numerical simulations show that in the presence of nonlinearity, dynamical localization is destroyed. Techniques of averaging and local derivatives on logarithmic scales indicates existence of the theoretically predicted spreading regimes of weak and strong chaos, as well as the crossover between them. This supports the idea that delocalization can be understood as a result of mode interactions inside the packet with those outside, as induced by the nonlinearity shift. In addition, the existence of a crossover might be associated with findings that the exponent of subdiffusive growth of the second moment can be temporarily close to $0.4$, which was obtained in earlier papers on the nonlinear quantum kicked rotor \cite{shepelyansky_delocalization_1993, garcia-mata_delocalization_2009}. Our simulations also confirm a novel lack of a self-trapping regime, which is not expected in this model due to an unbounded - yet circular - linear spectrum. We anticipate higher-dimensional generalized theoretical frameworks developed for nonlinear disordered chains \cite{flach_spreading_2010} to also be valid in the ``higher-dimensional'' models of kicked rotor systems \cite{lemarie_observation_2009,lemarie_critical_2010}.
\acknowledgments{The authors wish to thank T. V. Laptyeva, Ch. Skokos, D. O. Krimer, and P. Anghel-Vasilescu for group discussions. G. G. acknowledges support from 
the Ministry of Science, Serbia (Project III 45010).} 
\bibliographystyle{eplbib.bst}
\bibliography{QNKR}

\begin{thebibliography}{10}
\expandafter\ifx\csname url\endcsname\relax\def\url#1{\texttt{#1}}\fi

\bibitem{casati_stochastic_1979}
\Name{Casati G., Chirikov B., Izraelev F. \and Ford J.}
  \REVIEW{Lecture~Notes~in~Phys.}{93}{1979}{334}.

\bibitem{chirikov_universal_1979}
\Name{Chirikov B.} \REVIEW{Phys.~Rep.}{52}{1979}{263}.

\bibitem{fishman_chaos_1982}
\Name{Fishman S., Grempel D. \and Prange R.} \REVIEW{Phys.~Rev.~Lett.}{49}{1982}{509}.

\bibitem{grempel_quantum_1984}
\Name{Grempel D., Prange R. \and Fishman S.} \REVIEW{Phys.~Rev.~A}{29}{1984}{1639}.

\bibitem{bayfield_localization_1989}
\Name{Bayfield J., Casati G., Guarneri I. \and Sokol D.}
  \REVIEW{Phys.~Rev.~Lett.}{63}{1989}{364}.

\bibitem{bluemel_dynamical_1991}
\Name{Bl\"{u}mel R., Buchleitner A., Graham R., Sirko L., Smilansky U. \and
  Walther H.} \REVIEW{Phys.~Rev.~A}{44}{1991}{4521}.

\bibitem{ryu_high-order_2006}
\Name{Ryu C., Andersen M., Vaziri A., {d\textquoteright Arcy} M., Grossman J.,
  Helmerson K. \and Phillips W.} \REVIEW{Phys.~Rev.~Lett.}{96}{2006}{160403}.

\bibitem{behinaein_exploring_2006}
\Name{Behinaein G., Ramareddy V., Ahmadi P. \and Summy G.}
  \REVIEW{Phys.~Rev.~Lett.}{97}{2006}{244101}.

\bibitem{kanem_observation_2007}
\Name{Kanem J., Maneshi S., Partlow M., Spanner M. \and Steinberg A.}
  \REVIEW{Phys.~Rev.~Lett.}{98}{2007}{083004}.

\bibitem{shepelyansky_delocalization_1993}
\Name{Shepelyansky D.} \REVIEW{Phys.~Rev.~Lett.}{70}{1993}{1787}.

\bibitem{garcia-mata_delocalization_2009}
\Name{{Garc\'{\i}a-Mata} I. \and Shepelyansky D.} \REVIEW{Phys.~Rev.~E}{79}{2009}{026205}.

\bibitem{lemarie_observation_2009}
\Name{Lemari\'{e} G., Chab\'{e} J., Szriftgiser P., Garreau J., Gr\'{e}maud B.
  \and Delande D.} \REVIEW{Phys.~Rev.~A}{80}{2009}{043626}.

\bibitem{lemarie_critical_2010}
\Name{Lemari\'{e} G., Lignier H., Delande D., Szriftgiser P. \and Garreau J.}
  \REVIEW{Phys.~Rev.~Lett.}{105}{2010}{}.

\bibitem{flach_spreading_2010}
\Name{Flach S.} \REVIEW{Chem.~Phys.}{375}{2010}{548}.

\bibitem{krimer_statistics_2010}
\Name{Krimer D. \and Flach S.} \REVIEW{Phys.~Rev.~E}{82}{2010}{046221}.

\bibitem{laptyeva_crossover_2010}
\Name{Laptyeva T.~V., Bodyfelt J.~D., Krimer D.~O., Skokos C. \and Flach S.}
  \REVIEW{Europhys.~Lett.}{91}{2010}{30001}.

\bibitem{bodyfelt_nonlinear_2011}
\Name{Bodyfelt J.~D., Laptyeva T.~V., Skokos C., Krimer D.~O. \and Flach S.}
  \REVIEW{Phys.~Rev.~E}{84}{2011}{016205}.

\bibitem{izrailev_simple_1990}
\Name{Izrailev F.} \REVIEW{Phys.~Rep.}{196}{1990}{299}.

\bibitem{brenner_pseudo-randomness_1992}
\Name{Brenner N. \and Fishman S.} \REVIEW{Nonlinearity}{5}{1992}{211}.

\bibitem{rebuzzini_delocalized_2005}
\Name{Rebuzzini L., Wimberger S. \and Artuso R.} \REVIEW{Phys.~Rev.~E}{71}{2005}{036220}.

\bibitem{skokos_delocalization_2009}
\Name{Skokos C., Krimer D., Komineas S. \and Flach S.} \REVIEW{Phys.~Rev.~E}{79}{2009}{056211}.

\bibitem{kopidakis_absence_2008}
\Name{Kopidakis G., Komineas S., Flach S. \and Aubry S.}
  \REVIEW{Phys.~Rev.~Lett.}{100}{2008}{084103}.

\bibitem{cleveland_locally_1988}
\Name{Cleveland W.~S. \and Devlin S.~J.} \REVIEW{J.~Am.~Stat.~Assoc.}{83}{1988}{596}.

\bibitem{hoffman_numerical_1992}
\Name{Hoffman J.} \Book{Numerical methods for engineers and scientists}
  ({McGraw-Hill}, New York) 1992.

\end{thebibliography}
\end{document}